\documentstyle[aps,preprint]{revtex}
\begin{document}
\title{A Brief Review of Future Lepton-Hadron and Photon-Hadron Colliders}
\author{S. Sultansoy}
\address{Physics Dept., Faculty of Arts and Sciences, Gazi University, 06500,\\
Ankara, TURKEY \\
Institute of Physics, Academy of Sciences, H. Cavid Ave. 33, Baku,\\
AZERBAIJAN}
\author{A.K. \c{C}ift\c{c}i, E. Recepo\u{g}lu}
\address{Dept. of Physics, Faculty of Sciences, Ankara University, 06100\\
Tandogan, Ankara, TURKEY}
\author{\"{O}. Yava\c{s}}
\address{Dept. of Eng. Physics, Faculty of \ Engineering, Ankara University, 06100\\
Ankara, TURKEY}
\maketitle

\begin{abstract}
Options for future $lepton-proton,lepton-nucleus,\gamma -proton,\gamma
-nucleus$ and FEL$\gamma -nucleus$ colliders are discussed. In the spirit of
this content, we consider TESLA$\otimes $HERA, LEP$\otimes $LHC, $\mu -$ring$%
\otimes $TEVATRON, e$\otimes $RHIC, Linac$\otimes $LHC, $\sqrt{s}=3TeV$ $\mu
p$ and CLIC based colliders.
\end{abstract}

\section{\protect\bigskip Introduction}

It is known that lepton-hadron collisions have been playing a crucial role
in exploration of deep inside of Matter. For example, the quark-parton model
was originated from investigation of electron-nucleon scattering. HERA has
opened a new era in this field extending the kinematics region by two orders
both in high Q$^{2}$ and small $x$ compared to fixed target experiments.
However, the region of sufficiently small $x$ and simultaneously high Q$^{2}$
( $\geq $10 GeV$^{2}$), where saturation of parton densities should manifest
itself, is currently not achievable. It seems possible that eA option of
HERA will give opportunity to observe such phenomena. Then, the acceleration
of polarized protons in HERA could provide clear information on nucleon spin
origin.

The investigation of physics phenomena at extreme small $x$ but sufficiently
high Q$^{2}$ is very important for understanding the nature of strong
interactions at all levels from nucleus to partons. At the same time, the
results from lepton-hadron colliders are necessary for adequate
interpretation of physics at future hadron colliders. Today, linac-ring type
machines seem to be the main way to TeV scale in lepton-hadron collisions;
however, it is possible that in future $\mu p$ machines can be added
depending on solutions of principal issues of basic $\mu ^{+}\mu ^{-}$
colliders.

Construction of future lepton linacs tangentially to hadron rings (HERA,
Tevatron or LHC) will provide a number of additional opportunities to
investigate lepton-hadron and photon-hadron interactions at TeV scale (see
[1-3] and references therein). For example:

\begin{center}
TESLA$\otimes $HERA = TESLA$\oplus $HERA

$\oplus $TeV scale $ep$ collider

$\oplus $TeV scale $\gamma p$ collider

\ \ $\oplus eA$ collider \ \ \ \ \ \ \ \ \ \ \ \ \ \ \ 

$\ \ \oplus \gamma A$ collider \ \ \ \ \ \ \ \ \ \ \ \ \ \ \ 

$\oplus $FEL $\gamma $A collider. \ \ \ \ \ \ 
\end{center}

\bigskip It should be noted that $\gamma p$ and $\gamma A$ options are
unique features of linac-ring type machines and can not be realized at LEP$%
\otimes $LHC, while it comparable with TESLA$\otimes $HERA $ep$ and $eA$
options.

There are a number of reasons [4,5] favoring a superconducting linear
collider (such as TESLA) as a source of e-beam for linac-ring type
colliders. First of all, spacing between bunches in warm linacs, which is of
the order of ns, doesn't match with the bunch spacing in the HERA, TEVATRON
and LHC. Also the pulse length is much shorter than the ring circumference.
In the case of TESLA, which use standing wave cavities, one can use both
shoulders of linac in order to double electron beam energy, whereas in the
case of conventional linear colliders one can use only half of the machine,
because the travelling wave structures can accelerate only in one direction.

The most transparent expression for the luminosity of linac-ring type $ep$
colliders is [5]:

\[
L_{ep}=\frac{1}{4\pi }\cdot \frac{P_{e}}{E_{e}}\cdot \frac{n_{p}}{%
\varepsilon _{p}^{N}}\cdot \frac{\gamma _{p}}{\beta _{p}^{\ast }} 
\]
\bigskip for round, transversely matched beams. The lower limit on $\beta
_{p}^{\ast }$, which is given by proton bunch length, can be overcome by
applying a ''dynamic'' focusing scheme [6], where the proton bunch waist
travels with electron bunch during collision. In this scheme $\beta
_{p}^{\ast }$ is limited, in principle, by the electron bunch length, which
is two orders magnitude smaller. More conservatively, an upgrade of the
luminosity by a factor 3-4 may be possible.

Earlier, the idea of using high energy photon beams, obtained by Compton
backscattering of laser light off a beam of high energy electrons, was
considered for $\gamma e$ and $\gamma \gamma $ colliders (see [7] and
references therein). Then the same method was proposed for constructing $%
\gamma p$ colliders on the base of linac-ring type $ep$ machines in [8].
Rough estimations of the main parameters of $\gamma p$ collisions are given
in [9]. The dependence of these parameters on the distance z between
conversion region and collision point was analyzed in [10], where some
design problems were considered.

The aim of this brief review is to draw the attention of the HEP community
to future $lp,lA,\gamma p,\gamma A$ and FEL $\gamma A$ collider facilities.

\section{First stage: THERA, LEP$\otimes $LHC, $\protect\mu \otimes $%
TEVATRON and $e\otimes $RHIC}

\subsection{THERA}

Recently, the work on TESLA TDR has been finished, and TESLA$\otimes $HERA
based $ep,\gamma p,eA$ and $\gamma A$ colliders are included [11] into
TESLA\ project.

\subsubsection{$ep$ option}

Main parameters of TESLA$\otimes $HERA based on $ep$ collider are given in
Table I. It is seen that one has $L_{ep}=4.1\cdot 10^{30}$ cm$^{-2}$s$^{-1}$%
\ with $E_{e}=250$ GeV and $E_{p}=1$ TeV. Also two additional versions ($%
E_{e}=E_{p}=500$ GeV with $L_{ep}=2.5\cdot 10^{31}$ cm$^{-2}$s$^{-1}$ and $%
E_{e}=$ $E_{p}=800$ GeV with $L_{ep}=1.6\cdot 10^{31}$ cm$^{-2}$s$^{-1}$)
have been mentioned.

In principle, TESLA$\otimes $ HERA based ep collider will extend the HERA
kinematics region by an order in both Q$^{2}$ and $x$ and, therefore, the
parton saturation regime can be achieved. A brief account of some SM physics
topics (structure functions, hadronic final states, high Q$^{2}$ region
etc.)which can be searched in TESLA$\otimes $HERA $ep$ collider is presented
in [11]. The BSM search capacity of the machine will be defined by future
results from LHC. If the first family leptoquarks and/or leptogluons have
masses less than 1 TeV, they will be produced copiously (for couplings of
the order of $\alpha _{em}$). The indirect manifestation of new gauge bosons
may also be a matter of interest. In general, the physics search program of
the machine is a direct extension of the HERA search program.

\subsubsection{$\protect\gamma p$ option}

Referring to [10] for details let us note that $L_{\gamma p}\approx 2L_{ep}$%
\ at z=0 (for $\gamma $ options we use $\varepsilon _{e}=10^{-6\text{ }}$m$%
), $ where $z$ is distance between conversion region and collision point.
Then, as one can see from Fig. 1, luminosity slowly decreases with the
increasing z (factor $\sim $1/2 at z=10 m) and opposite helicity values for
laser and electron beams are advantageous (see Fig. 2). Additionally, a
better monochromatization of high-energy photons seen by proton bunch can be
achieved by increasing the distance $z$ (Fig. 3).

The scheme with non-zero crossing angle and electron beam deflection
considered in [10] for $\gamma p$ option lead to problems due to intensive
synchrotron radiation of bending electrons and necessity to avoid the
passing of electron beam from the proton beam focusing quadrupoles.
Alternatively, one can assume head-on-collisions (see above) and exclude
deflection of electrons after conversion. In this case residual electron
beam will collide with proton beam together with high-energy $\gamma $ beam,
but because of larger cross-section of $\gamma p$ interaction the background
resulting from ep collisions may be neglected. The problem of over-focussing
of the electron beam by the strong proton-low-$\beta $ quadrupoles is solved
using the fact of smallness of the emittance of the TESLA electron beam. For
this reason the divergence of the electron beam after conversion will be
dominated by the kinematics of the Compton backscattering. In the case of
250 (800) GeV electron beam the maximum value of scattering angle is 4 (1.5)
micro-radians. Therefore, the electron beam transverse size will be 100
(37.5) $\mu m$ at the distance of 25 m from conversion region and the
focusing quadrupoles for proton beam have negligible influence on the
residual electrons. On the other hand, in the scheme with deflection there
is no restriction on n$_{e}$ from $\Delta Q_{p}$, therefore, larger $n_{e}$
and bunch spacing may be preferable. All these topics need a further
research.

Concerning the experimental aspects, very forward detector in $\gamma $-beam
direction will be very useful for investigation of small $x_{g}$ region due
to registration of charmed and beauty hadrons produced via $\gamma
g\longrightarrow \overline{Q}Q$ sub-process.

There are a number of papers (see [2] and refs. therein), devoted to physics
at $\gamma p$ colliders. Concerning the BSM physics, $\gamma p$ option of
THERA does not promise essential results with possible exclusions of the
first family excited quarks (if their masses are less than 1 TeV) and
associate production of gaugino and first family squarks (if the sum of
their masses are less than 0.5 TeV). The photo-production of W and Z bosons
may be also the matter of interest for investigation of the their anomalous
couplings. However, $\overline{c}c$ and $\overline{b}b$\ pairs will be
copiously photo-produced at $x_{g}$ of order of 10$^{-5}$ and 10$^{-4}$,
respectively, and saturation of gluons should manifest itself. Then, there
are a number of different photo-production processes (including di-jets
etc.) which can be investigated at $\gamma p$\ colliders.

\subsubsection{$eA$ option}

The main limitation for this option comes from fast emittance growth due to
intra-beam scattering. In this case, the use of flat nucleus beams seems to
be more advantageous (as in the case of ep option [12]) because of
luminosity lifetime increase of few times. Nevertheless, sufficiently high
luminosity can be achieved at least for light nuclei. For example, $L_{eC}=$%
10$^{29}$ cm$^{-2}$s$^{-1}$ for collisions of 250 GeV energy electrons beam
(Table I) and Carbon beam with and \ n$_{C}=2.5\cdot 10^{9}$ (rest of
parameters as in Table I). This value corresponds to $L^{int}\cdot A\approx
10$ pb$^{-1\text{ }}$per working year ($10^{7}s$), needed from the physics
point of view [13,14].

As mentioned above, the large charge density of nucleus bunch results in
strong intra-beam scattering effects and lead to essential reduction of
luminosity lifetime ($\approx $ 1 h for C beam at HERA). There are two
possible solutions of this problem for TESLA$\otimes $HERA. Firstly, one
could consider the possibility to re-fill nucleus ring at appropriate rate
with necessary modifications of the filling time etc. Alternatively, an
effective method of cooling of nucleus beam in main ring should be applied,
especially for heavy nuclei. For example, electron cooling of nucleus beams
suggested in [15] for $eA$ option of HERA can be used for TESLA$\otimes $%
HERA, also.

The physics search program of the machine is the direct extension of that
for eA option of HERA (see Chapter titled ''Light and Heavy Nuclei in HERA''
in [16]).

\subsubsection{$\protect\gamma A$ option}

In our opinion this is the most promising option of TESLA$\otimes $HERA
complex, because it will give unique opportunity to investigate small $x_{g}$
region in nuclear medium. Indeed, due to the advantage of the real $\gamma $%
\ spectrum, heavy quarks will be produced via $\gamma g$ fusion at
characteristic

\[
x_{g}\approx \frac{4m_{c(b)}^{2}}{0.9\times (Z/A)\times s_{ep}} 
\]
which is approximately $\left( 2\div 3\right) \cdot 10^{-5}$ for charmed
hadrons.

As in the previous option, sufficiently high luminosity can be achieved at
least for light nuclei. The scheme with deflection of electron beam after
conversion is preferable because it will give opportunity to avoid
limitations from $\Delta Q_{A}$, especially for heavy nuclei. The dependence
of luminosity on the distance between conversion region and interaction
point for HERA based $\gamma C$\ collider is presented in Figure 4 and $%
L_{\gamma C}=0.6\cdot 10^{29}$ cm$^{-2}$s$^{-1}$ at $z=5$ m. Let us remind
you that an upgrade of the luminosity by a factor 3-4 may be possible by
applying a ''dynamic'' focusing scheme. Further increase on luminosity will
be possible with the cooling of nucleus beam in the main ring. Finally, very
forward detector in $\gamma $-beam direction will be very useful for
investigation of small $x_{g}$\ region due to registration of charmed and
beauty hadrons.

Let us finish this section by quoting the paragraph, written for eA option
of the TESLA$\otimes $HERA complex but more applicable for $\gamma A$\
option, from [13]:

''Extension of the $x$-range by two orders of magnitude at TESLA-HERA
collider would correspond to an increase of the gluon densities by a factor
of 3 for $Q^{2}=10$ GeV$^{2}$. It will definitely bring quark interactions
at this scale into the region where DGLAP will break down. For the
gluon-induced interactions it would allow the exploration of a non-DGLAP
hard dynamics over two orders of magnitude in $x$ in the kinematics where $%
\alpha _{s}$ is small while the fluctuations of parton densities are large.''

\subsubsection{FEL $\protect\gamma A$ option}

Colliding of TESLA FEL beam with nucleus bunches from HERA may give a unique
possibility to investigate ''old'' nuclear phenomena in rather unusual
conditions. The main idea is very simple [2,17]: ultra-relativistic ions
will see laser photons with energy $\omega _{o}$\ as a beam of photons with
energy 2$\gamma _{A}\omega _{o}$\ , where $\gamma _{A}$\ is the Lorentz
factor of the ion beam. Moreover, since the accelerated nuclei are fully
ionized, we will be free from possible background induced by low-shell
electrons. For HERA $\gamma _{A}=(Z/A)\gamma _{p}\approx 980(Z/A)$ ,
therefore, the region 0.1$\div $10 MeV, which is matter of interest for
nuclear spectroscopy, corresponds to 0.1$\div $10 keV lasers, which coincide
with the energy region of TESLA FEL.

The excited nucleus will return to the ground state at a distance $l=\gamma
_{A}.\tau _{A}.c$ from the collision point, where $\tau _{A}$ is the
lifetime of the exited state in the nucleus rest frame and $c$ is the speed
of light. For example, one has $l=4$\ mm for 4438 keV excitation of $^{12}$%
C. Therefore, the detector should be placed close to the collision region.
The MeV energy photons emitted in the rest frame of the nucleus will be seen
in the detector as high-energy photons with energies up to GeV region.

The huge number of expected events ($\sim $10$^{10}$ per day for 4438 keV
excitation of $^{12}$C) and small energy spread of colliding beams ($\leq
10^{-3}$ for both nucleus and FEL beams) will give opportunity to scan an
interesting region with $\sim $1 keV accuracy.

\subsection{ LEP$\otimes $LHC}

The interest in this collider, which was widely discussed [18,19] at earlier
stages of LHC proposal, has renewed recently [20]:

''We consider the LHC $e^{\pm }p\ $option to be already part of the LHC
programme. The availability of $e^{\pm }p$\ collisions at an energy roughly
four times that provided currently by HERA would allow studies of quark
structure down to a size of about 10$^{-17}$ cm... The discovery of the
quark substructure could explain the Problem of Flavour, or one might
discover leptoquarks or squarks as resonances in the direct channel...''

\subsubsection{$ep$ option}

The recent set of parameters is given in a report [21] prepared at the
request of the CERN Scientific Policy Committee. With these parameters the
estimated luminosity is $L_{ep}=1.2\cdot 10^{32}$ cm$^{-2}$s$^{-1}$ and
exceeds that of the HERA$\otimes $TESLA based $ep$\ collider. However, the
latter has the advantage in kinematics because of comparable values of
energies of colliding particles. Moreover, $\gamma p$\ collider can not be
constructed on the base of LEP$\otimes $LHC (for reasons see [10]).

\subsubsection{$eA$ option}

An estimation of the luminosity for $e^{\pm }Pb$ collisions given in [21]
seems to be over-optimistic because of unacceptable value of $\Delta Q_{Pb}$%
\ , which is $\sim 70\cdot 10^{-3}$ for proposed set of parameters. The one
order lower value $L_{e-Pb}\approx 10^{28}$ cm$^{-2}$s$^{-1}$ is more
realistic. However, situation may be different for light nuclei. Again the
main advantage of the TESLA$\otimes $HERA complex in comparison with LEP$%
\otimes $LHC is $\gamma A$\ option.

\subsection{$\protect\mu \otimes $TEVATRON}

If the main problems (high $\mu -$production rate, fast cooling of $\mu -$%
beam etc.) facing $\mu ^{+}\mu ^{-}$\ proposals are successfully solved, it
will also give an opportunity to construct $\mu p$ colliders. Today only
very rough estimations of the parameters of these machines can be made. Two
sets of parameters for the collider with two rings, TEVATRON and 200 GeV
muon ring, are considered in [22]. In our opinion, luminosity values
presented in [22] namely, $L_{\mu p}=1.3\cdot 10^{31}$cm$^{-2}$s$^{-1}$ are
over-optimitic. For comparison, recent set of 200 GeV muon beam parameters
[23] leads to estimation $L_{\mu p}=1.7\cdot 10^{33}$ cm$^{-2}$s$^{-1}$ (for
details see review [3]). Physics search program of this machine is similar
to that of the ep option of the TESLA$\otimes $ HERA complex.

\subsection{e$\otimes $RHIC}

Recently, an addition of 10 GeV electron ring or linac to RHIC\ ring is
discussed in order to investigate lepton-nucleus interactions [24].
Luminosity of electron-gold collision for the first case is expected as $%
6.4(45)\cdot 10^{30}$ cm$^{-2}$s$^{-1}$ with 360 (2520) bunches in each
beam. For the second case expected luminosity is $5.6\cdot 10^{30}$ cm$^{-2}$%
s$^{-1}.$ However, in order to achieve these relatively high values an
electron cooling of nucleus beam is needed.

Beceuse of low value of the electron energy, $\gamma $ options with Compton
backscattering photons are not sufficiently advantageous. On the other side,
FEL $\gamma A$ options are the matter of interest if the TESLA-like
accelerator is chosen as an electron linac. Parameters of the FEL $\gamma $%
-Th collisions for e-linac$\otimes $RHIC are estimated in [25].

\section{Second stage: Linac$\otimes $LHC, $\surd s=3$ TeV $\protect\mu p$
and CLIC Based Colliders}

\subsection{Linac$\otimes $LHC}

The center-of-mass energies which will be achieved at different options of
this machine [26,27] are an order larger than those at HERA are and $\sim $3
times larger than the energy region of TESLA$\otimes $HERA, LEP$\otimes $LHC
and $\mu \otimes $TEVATRON. Following [4,5] below we consider electron linac
with $P_{e}$= 60 MW and upgraded proton beam from LHC (Table II).

\subsubsection{$ep$ option}

According to parameters given in Table II, center-of-mass energy and
luminosity for this option are $\sqrt{s}=5.29$ TeV and $L_{ep}=8\cdot
10^{31} $ cm$^{-2}$s$^{-1}$, respectively, and additional factor 3-4 can be
provided by the ''dynamic'' focusing scheme [6]. Further increasing will
require cooling at injector stages.

This machine, which will extend both the $Q^{2}$-range and $\ x$-range by
more than two order of magnitude comparing to those explored by HERA, has a
strong potential for both SM and BSM research.

\subsubsection{$\protect\gamma p$ option}

The advantage in spectrum of back-scattered photons (for details see ref.
[7-10]) and sufficiently high luminosity ($L_{\gamma p}>10^{32}$ cm$^{-2}$s$%
^{-1}$) will clearly manifest itself in a search for different phenomena.
For example, thousands di-jets with $p_{t}>500$\ GeV and hundreds thousands
single W bosons will be produced, hundred millions of \ $\overline{b}b$- and
\ $\overline{c}c$- pairs will give opportunity to explore the region of
extremely small x$_{g}$ etc [2].

In Fig. 5, the dependence of luminosity on the distance $z$ between
interaction point (IP) and conversion region (CR) is plotted (for $\gamma $
options we use $\varepsilon _{e}=10^{-6}$ m and $\beta _{x,y}^{e}=0.1$ m).
In Fig. 6, we plot luminosity distribution as a function of $\gamma p$
invariant mass $W_{\gamma p}=2\sqrt{E_{\gamma }E_{p}}$ at $z=10$ m. In Fig.
7, this distribution is given for choice of $\lambda _{e}=0.8$ and $\lambda
_{0}=-1$ at three different values of the distance between IP and CR.

\subsubsection{$eA$ option}

In the case of LHC nucleus beam IBS effects in main ring are not crucial
because of larger value of $\gamma _{A}$\ . The main principal limitation
for heavy nuclei coming from beam-beam tune shift may be weakened using flat
beams at collision point. Rough estimations show that $L_{eA}\cdot A>10^{31}$
cm$^{-2}$s$^{-1}$ can be achieved at least for light and medium nuclei [26].

\subsubsection{$\protect\gamma A$ option}

Limitation on luminosity due to beam-beam tune shift is removed in the
scheme with deflection of electron beam after conversion. In Fig. 8, the
dipendence of luminosity on $z$ is plotted for linac-LHC based $\gamma -Pb$
collider, where we use $10^{8}$ for number of lead nuclei per bunch and the
rest of parameters are as in Table II.

The physics search potential of this option, as well as that of previous
three options, needs more investigations from both particle and nuclear
physics viewpoints.

\subsubsection{FEL $\protect\gamma A$ option}

Due to a larger $\gamma _{A}$\ the requirement on wavelength of the FEL
photons is weaker than in the case of TESLA$\otimes $HERA based FEL $\gamma
A $ collider. Therefore, the possibility of constructing a special FEL for
this option may be matter of interest. In any case the realization of FEL $%
\gamma A$ colliders depends on the interest of ''traditional'' nuclear
physics community. The potential of these machine for investigations of Sm
excitations is considered in [28].

\subsection{$\surd s=3$ TeV $\protect\mu p$}

The possible $\mu p$ collider with $\sqrt{s}=4$ TeV in the framework of $\mu
^{+}\mu ^{-}$ project was discussed in [29] and again over-estimated value
of luminosity, namely, $L_{\mu p}=3\cdot 10^{35}$ cm$^{-2}$s$^{-1}$, was
considered. Using recent set of parameters [25] for high-energy muon
collider with $\sqrt{s}=3$ TeV, one can easily estimate possible parameters
of $\mu p$ collisions from:

\[
L_{\mu p}=\frac{n_{p}}{n_{\mu }}\cdot \frac{\beta _{\mu }^{\ast }}{\beta
_{p}^{\ast }}\cdot \frac{m_{\mu }}{m_{p}}\cdot \frac{\varepsilon _{\mu }^{N}%
}{\varepsilon _{p}^{N}}\cdot L_{\mu ^{+}\mu ^{-}} 
\]
With $n_{p}=n_{\mu }=2\cdot 10^{12}$ we obtain $L_{\mu p}=10^{32}$ cm$^{-2}$s%
$^{-1}$ (for details, see [3]).

This machine is comparable with the ep option of the Linac$\otimes $LHC.

\subsection{CLIC Based Lepton-Hadron and Photon-Hadron Colliders}

The CLIC [30], an electron-positron collider with $\sqrt{s}=3$ TeV and $%
L_{ee}=10^{35}$ cm$^{-2}$s$^{-1},$ is considered as one of the future
options for post-LHC era at CERN. If a $\sim 5$ GeV proton ring is added at
the beginning of positron shoulder of CLIC an opportunity to construct an $%
ep $ collider with $\sqrt{s}=3$ TeV and $\gamma p$ collider with $\sqrt{s}%
\approx 2.8$ TeV will appear. In order to coincide with CLIC parameters we
need a proton ring with $\sim 50$ m circumference and repetition frequency \ 
%TCIMACRO{\TEXTsymbol{>}}%
%BeginExpansion
\mbox{$>$}%
%EndExpansion
100 Hz. The luminosity of $ep$ collisions can be roughly estimated as $%
L_{ep}\approx (m_{e}/m_{p})\cdot L_{ee}\approx 5\cdot 10^{31}$cm$^{-2}$s$%
^{-1}$ if the parameters of proton beam coincide with those of CLIC positron
beam. In principle, this value can be increased by an order of magnitude due
to improvement of proton beam parameters. The work on the subject, as well
as on $\gamma p,eA,\gamma A$ and FEL $\gamma A$ options is under progress. \ 

\section{Conclusion}

It seems that neither HERA nor LHC$\otimes $ LEP will be the end points for
lepton-hadron colliders. Linac-ring type $ep$ machines and possibly $\mu p$
colliders will give an opportunity to go far in this direction (see Table
III). However, more activity is needed both in accelerator (further
exploration of ''dynamic'' focusing scheme, a search for effective cooling
methods etc.) and physics search program aspects.

{\Large Acknowledgements}

We would like to express our gratitude to DESY Directorate for invitations
and hospitality. We are grateful to D. Barber, E. Boos, V. Borodulin, R.
Brinkmann, O. Cakir, A. Celikel, L. Frankfurt, I. Ginzburg, P. Handel, M.
Klein, M. Leenen, C. Niebuhr, V. Serbo, M. Strikman, V. Telnov, D. Trines,
G. A. Voss, A. Wagner, N. Walker and F. Willeke for useful and stimulating
discussions. Our work on the subject was strongly encouraged by the support
of Professor B. Wiik, who personally visited Ankara in 1996 and signed the
Collaboration Agreement between DESY and Ankara University.

\bigskip \newpage

\bigskip \bigskip 
%TCIMACRO{
%\TeXButton{Table1}{\begin{table}[tbp] \centering%
%}}%
%BeginExpansion
\begin{table}[tbp] \centering%
%
%EndExpansion
\caption{ Main parameters of an ep collider based on HERA and
TESLA\label{key}}

\begin{tabular}{|l|l|}
\hline
\multicolumn{2}{|l|}{\ \ \ \ \ \ \ \ \ \ \ \ \ \ \ \ \ \ \ \ \ \ \ Electron
beam parameters} \\ \hline
Electron energy & E$_{e}=250$ GeV \\ \hline
Number of electrons per bunch & N$_{e}=2\cdot 10^{10}$ \\ \hline
Bunch length & $\sigma _{ze}=0.3$ mm \\ \hline
Invariant emittance & $\varepsilon _{e}=$100$\cdot 10^{-6}$ m \\ \hline
Beta function at IP & $\beta _{x,y}^{e}=0.5$ m \\ \hline
Bunch spacing & 211.37 ns \\ \hline
Number of bunches & 5264 \\ \hline
Repetition rate & 5 Hz \\ \hline
Beam power & 22.6 MW \\ \hline
\multicolumn{2}{|l|}{\ \ \ \ \ \ \ \ \ \ \ \ \ \ \ \ \ \ \ \ \ \ \ \ \
Proton beam parameters} \\ \hline
Proton energy & E$_{p}=1$ GeV \\ \hline
Number of protons per bunch & N$_{p}=10^{11}$ \\ \hline
Number of bunches & 94 \\ \hline
Bunch length & 10 cm \\ \hline
Beta function at IP & $\beta _{x,y}^{p}=0.1$ m \\ \hline
Normalized emittance & $\varepsilon _{p}=$1$\cdot 10^{-6}$ m \\ \hline
IBS growth time & $\tau _{s}=2.88$ h$,\tau _{x}=2$ h \\ \hline
\end{tabular}
%TCIMACRO{
%\TeXButton{E}{\end{table}%
%}}%
%BeginExpansion
\end{table}%
%
%EndExpansion
\newpage 

%TCIMACRO{
%\TeXButton{Table2}{\begin{table}[tbp] \centering%
%}}%
%BeginExpansion
\begin{table}[tbp] \centering%
%
%EndExpansion
\caption{Main parameters of an $ep$ collider based on LHC and
e-linac\label{key}} 
\begin{tabular}{|l|l|}
\hline
\multicolumn{2}{|l|}{\ \ \ \ \ \ \ \ \ \ \ \ \ \ \ \ \ \ \ \ \ \ \ \ \ \
Electron beam parameters} \\ \hline
Electron energy & E$_{e}=1$ TeV \\ \hline
Number of electrons per bunch & N$_{e}=7\cdot 10^{9}$ \\ \hline
Bunch length & $\sigma _{ze}=1$ mm \\ \hline
Invariant emittance & $\varepsilon _{e}=$10$\cdot 10^{-6}$ m \\ \hline
Beta function at IP & $\beta _{x,y}^{e}=0.2$ m \\ \hline
Bunch spacing & 100 ns \\ \hline
Number of bunches & 5000 \\ \hline
Repetition rate & 10 Hz \\ \hline
Beam power & 56 MW \\ \hline
\multicolumn{2}{|l|}{\ \ \ \ \ \ \ \ \ \ \ \ \ \ \ \ \ \ \ \ \ \ \ \ \ \
Proton beam parameters} \\ \hline
Proton energy & E$_{p}=7$ GeV \\ \hline
Number of protons per bunch & N$_{p}=4\cdot 10^{11}$ \\ \hline
Number of bunches & 700 \\ \hline
Bunch length & 7.5 cm \\ \hline
Beta function at IP & $\beta _{x,y}^{p}=0.1$ m \\ \hline
Normalized emittance & $\varepsilon _{p}=$0.8$\cdot 10^{-6}$ m \\ \hline
IBS growth time & $\tau _{s}=2.5$ h$,\tau _{x}=5.2$ h \\ \hline
\end{tabular}
%TCIMACRO{
%\TeXButton{E}{\end{table}%
%}}%
%BeginExpansion
\end{table}%
%
%EndExpansion
\newpage 
%TCIMACRO{
%\TeXButton{Table3}{\begin{table}[tbp] \centering%
%}}%
%BeginExpansion
\begin{table}[tbp] \centering%
%
%EndExpansion
\caption{Future lepton-hadron colliders:\label{key}}

a) First stage (2010-2015)

\begin{tabular}{|l|l|l|l|l|}
\hline
& TESLA$\otimes $HERA & LEP$\otimes $LHC & $\mu \otimes $TEVATRON & e$%
\otimes $RHIC \\ \hline
$\sqrt{s},$ TeV & 1.0$\rightarrow $1.6 & 1.37 & 0.89 & 0.1 \\ \hline
$E_{l},$ TeV & 0.25$\rightarrow $0.8 & 0.0673 & 0.2 & 0.01 \\ \hline
$E_{p},$ TeV & 1 & 7 & 1 & 0.25 \\ \hline
$L,$ 10$^{31}$ cm$^{-2}$s$^{-1}$ & 1-10 & 12 & 1-10 & 46 \\ \hline
Main limitations & $P_{e,}\varepsilon _{p},\beta _{p}^{\ast }$ & $\Delta
Q_{e},\Delta Q_{p}$ & $n_{\mu },\varepsilon _{\mu },\Delta Q_{p},\varepsilon
_{p},\beta _{p}^{\ast }$ &  \\ \hline
Additional options & $eA,\gamma p,\gamma A,$ $FEL\gamma A$ & $eA$ & $\mu
A(?) $ & $eA,FEL\gamma A$ \\ \hline
\end{tabular}

b) Second stage (2015-2020) and third ($>$2020) stages

\begin{tabular}{|l|l|l|l|}
\hline
& Linac$\otimes $LHC & $\mu p$ & CLIC based \\ \hline
$\sqrt{s},$ TeV & 5.29 & 3 & 3 \\ \hline
$E_{l},$ TeV & 1 & 1.5 & 1.5 \\ \hline
$E_{p},$ TeV & 7 & 1.5 & 1.5 \\ \hline
$L,10^{31}$ cm$^{-2}$s$^{-1}$ & 10-100 & 10-100 & 10 \\ \hline
Options & $eA,\gamma p,\gamma A,$ $FEL\gamma A$ & $\mu A(?)$ & $eA,\gamma
p,\gamma A,$ $FEL\gamma A$ \\ \hline
\end{tabular}
%TCIMACRO{
%\TeXButton{E}{\end{table}%
%}}%
%BeginExpansion
\end{table}%
%
%EndExpansion

\bigskip \newpage

FIGURE CAPTIONS

\bigskip

Fig. 1. Luminosity dependence on distance $z$\ for the TESLA$\otimes $HERA
based $\gamma p$ collider.

\bigskip

Fig. 2. Luminosity distrubition as a function of $\gamma p$ invariant mass
at $z=10$\ m for the TESLA$\otimes $HERA based $\gamma p$ collider.

\bigskip

Fig. 3. Luminosity distrubition as a function of $\gamma p$ invariant mass
for $\lambda _{e}\lambda _{0}=-0.8$ \ at three different values of distance $%
z$ for the TESLA$\otimes $HERA based $\gamma p$ collider.

\bigskip

Fig. 4. Luminosity dependence on distance for the TESLA$\otimes $HERA based $%
\gamma C$ collider.

\bigskip

Fig. 5. Luminosity dependence on distance $z$\ for the Linac$\otimes $LHC
based $\gamma p$ collider.

\bigskip

Fig. 6. Luminosity distrubition as a function of $\gamma p$ invariant mass
at $z=10$\ m for the Linac$\otimes $LHC based $\gamma p$ collider.

\bigskip

Fig. 7. Luminosity distrubition as a function of $\gamma p$ invariant mass
for $\lambda _{e}\lambda _{0}=-0.8$ \ at three different values of distance $%
z$ for the Linac$\otimes $LHC based $\gamma p$ collider.

\bigskip

Fig. 8. Luminosity dependence on distance $z$ for the Linac$\otimes $LHC
based $\gamma Pb$ collider.


\begin{references}
\bibitem{1}  \bigskip R. Brinkmann et al., DESY 97-239 (1997); e-Print
Archive: physics/9712023.

\bibitem{2}  S. Sultansoy, Turkish J. Phys. 22 (1998) 575; e-Print Archive:
hep-ex/0007043.

\bibitem{3}  S. Sultansoy, DESY-99-159 (1999); e-Print Archive:
hep-ph/9911417.

\bibitem{4}  M. Tigner, B. Wiik and F. Willeke, Proceedings of the 1991 IEEE
Particle Accelerators Conference (6-9 May 1991, San Francisco, California),
vol. 5, p. 2910.

\bibitem{5}  B. H. Wiik, Proceedings of the International Europhysics
Conference on High Energy Physics (22-28 July 1993, Marseille, France), p.
739.

\bibitem{6}  R. Brinkmann and M. Dohlus, DESY-M-95-11 (1995).

\bibitem{7}  V.I. Telnov, Nucl. Instrum. Meth. A294 (1990) 72.

\bibitem{8}  S.I. Alekhin et al., Int. J. Mod. Phys. A6 (1991) 21.

\bibitem{9}  S.F. Sultanov, IC/89/409, Trieste (1989).

\bibitem{10}  A.K. Ciftci, S. Sultansoy, S. Turkoz and O. Yavas, Nucl.
Instrum. Meth. A365 (1995) 317.

\bibitem{11}  http://www.ifh.de/thera

\bibitem{12}  R. Brinkmann, Turkish J. Phys. 22 (1998) 661.

\bibitem{13}  L. Frankfurt and M. Strikman, Nucl. Phys. Proc. Suppl. 79
(1999) 671.

\bibitem{14}  M. Strikman, private communication.

\bibitem{15}  M. Gentner et al., Nucl. Instrum. Meth. A424 (1999) 277.

\bibitem{16}  G. Ingelman, A. De Roeck and R. Klanner (Eds.), Proceedings of
the workshop on Future Physics at HERA, DESY (September 1996).

\bibitem{17}  H. Aktas et al., Nucl. Instrum. Meth. A428 (1999) 271.

\bibitem{18}  W. Bartel, CERN 87-07, Vol. I, 303 (1987).

\bibitem{19}  A. Verdier, CERN 90-10, Vol. III, 820 (1990).

\bibitem{20}  J. Ellis, E. Keil and G. Rolandi, CERN-EP/98-03, CERN-SL
98-004 (AP), CERN-TH/98-33 (1998).

\bibitem{21}  E. Keil, LHC Project Report 93 (1997).

\bibitem{22}  V.D. Shiltsev, FERMILAB-Conf-97/114 (1997); FERMILAB-TM-1969
(1996).

\bibitem{23}  C.M. Ankenbrandt et al. (Muon Collider Collaboration), Phys.
Rev. SP-AB 2 (1999) 081001.

\bibitem{24}  I. Ben-Zvi, J. Kewish, J. Murphy and S. Peggs, C-A/AP/14
(2000).

\bibitem{25}  H.Koru, A. Ozcan, S. Sultansoy and B. Sarer, e-Print archive:
nucl-ex/0106010.

\bibitem{26}  O. Yavas, A. K. \c{C}ift\c{c}i and S. Sultansoy, Proceedings
of the 7 th Europan Particle Accelerator Conference (26-30 Jun 2000, Vienna,
Austria), p.391.

\bibitem{27}  A. K. \c{C}ift\c{c}i, S. Sultansoy and O. Yavas, e-Print
Archive: hep-ex/0006030.

\bibitem{28}  E. Guliyev, A.A. Kuliev, S. Sultansoy and O. Yavas, e-Print
Archive: nucl-ex/0105022.

\bibitem{29}  I.F. Ginzburg, Turkish J. Phys. 22 (1998) 607.

\bibitem{30}  http://cern.web.cern.ch/CERN/Divisions/PS/CLIC/Welcome.html
\end{references}
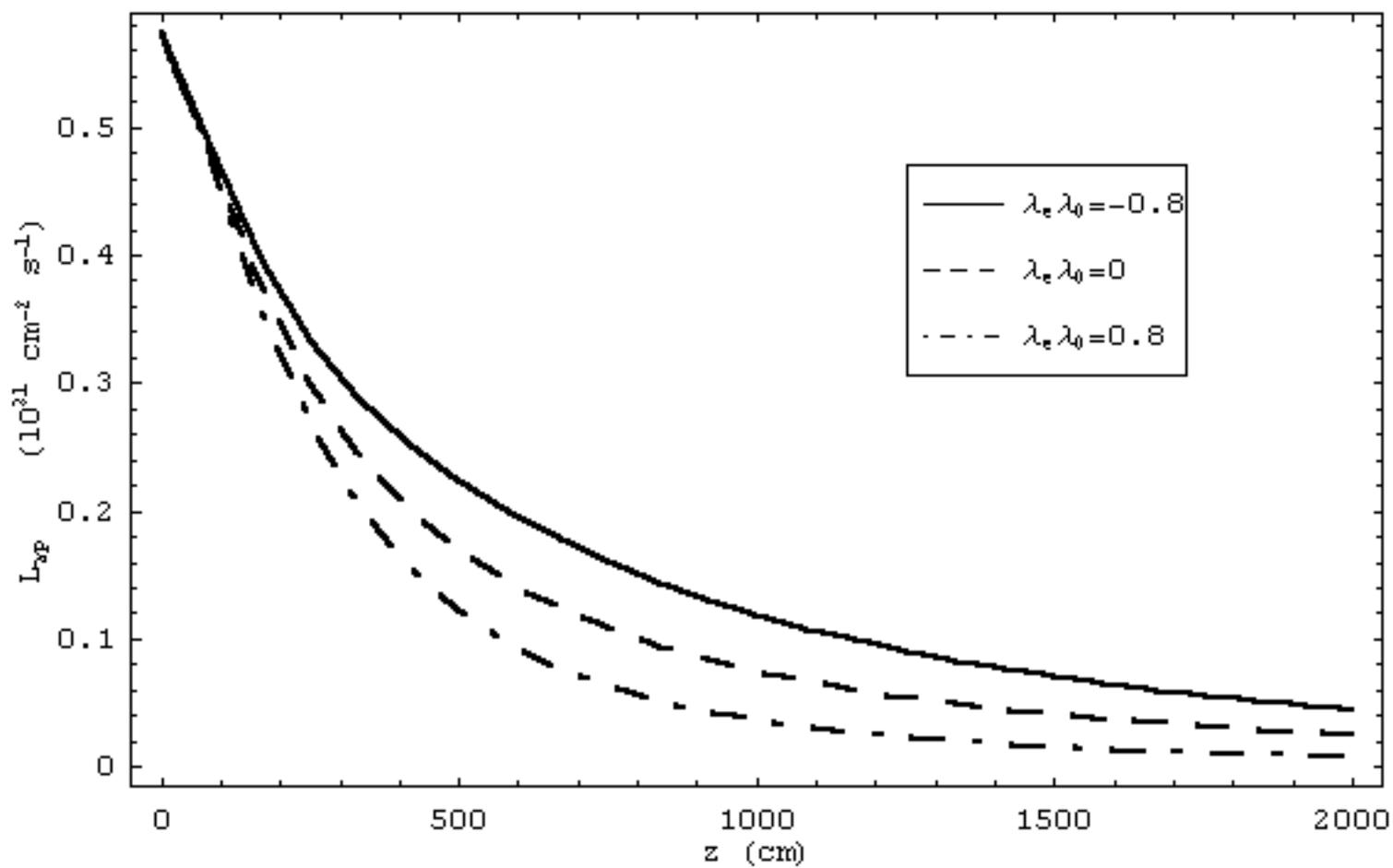
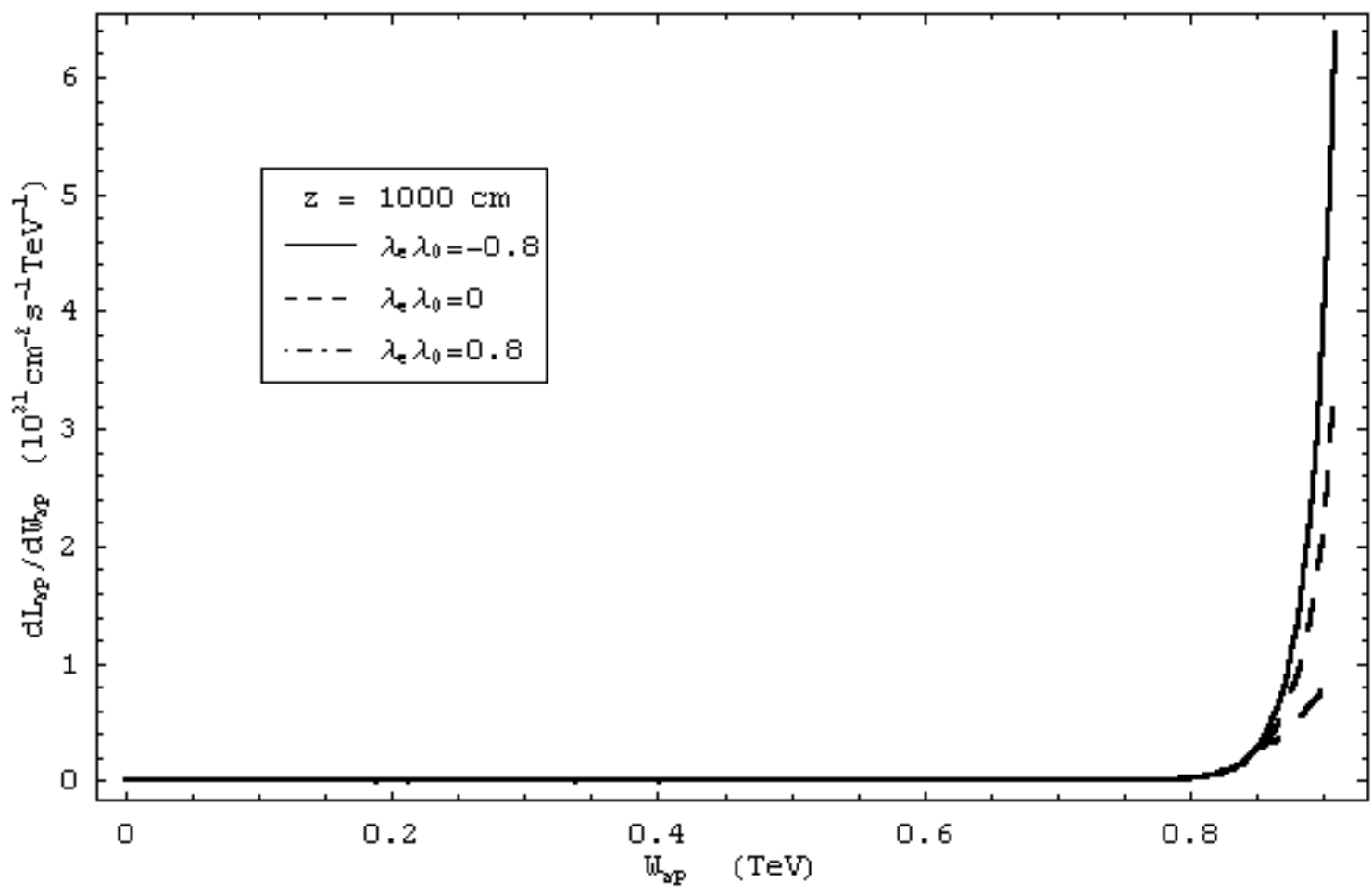
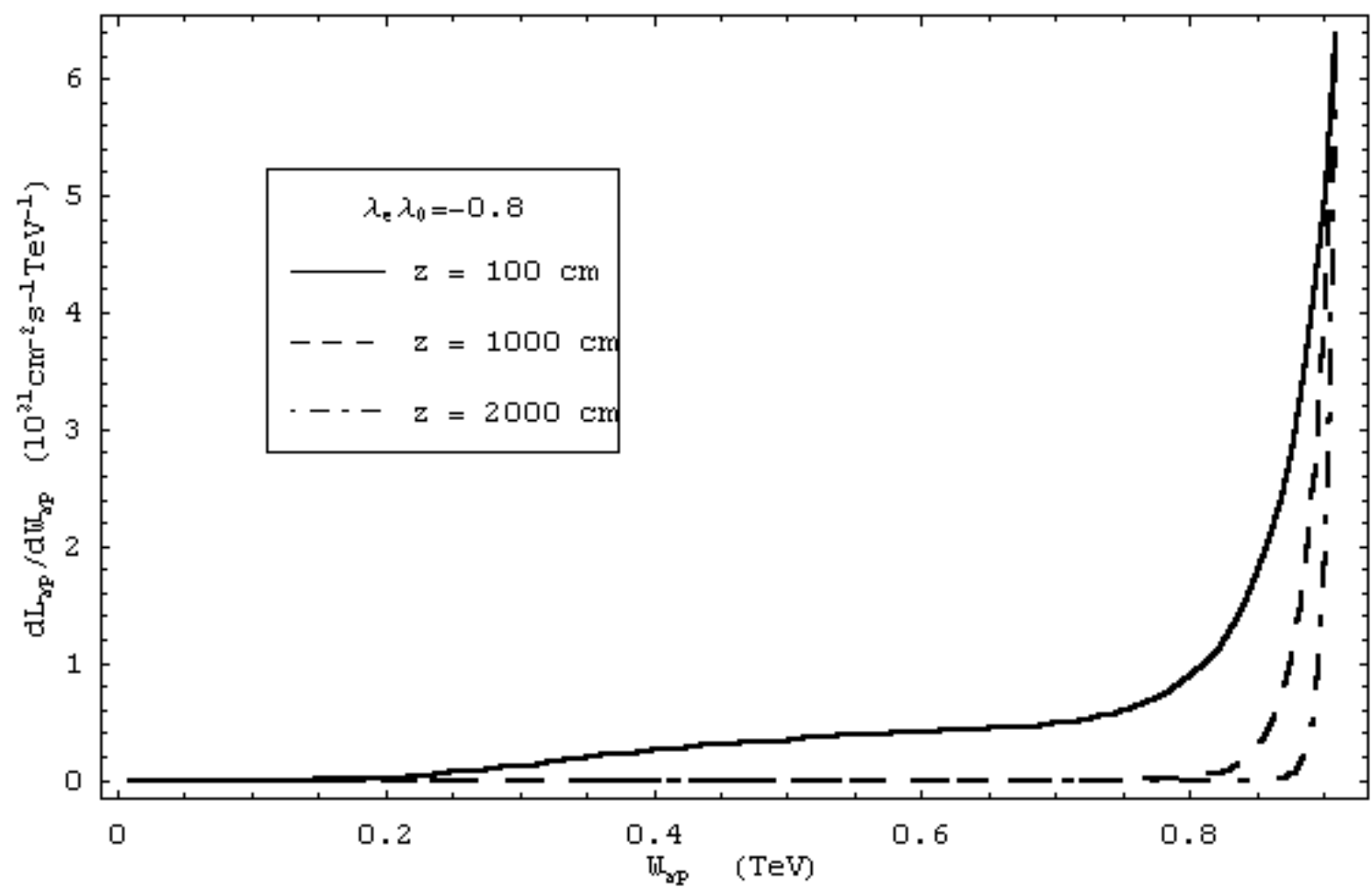
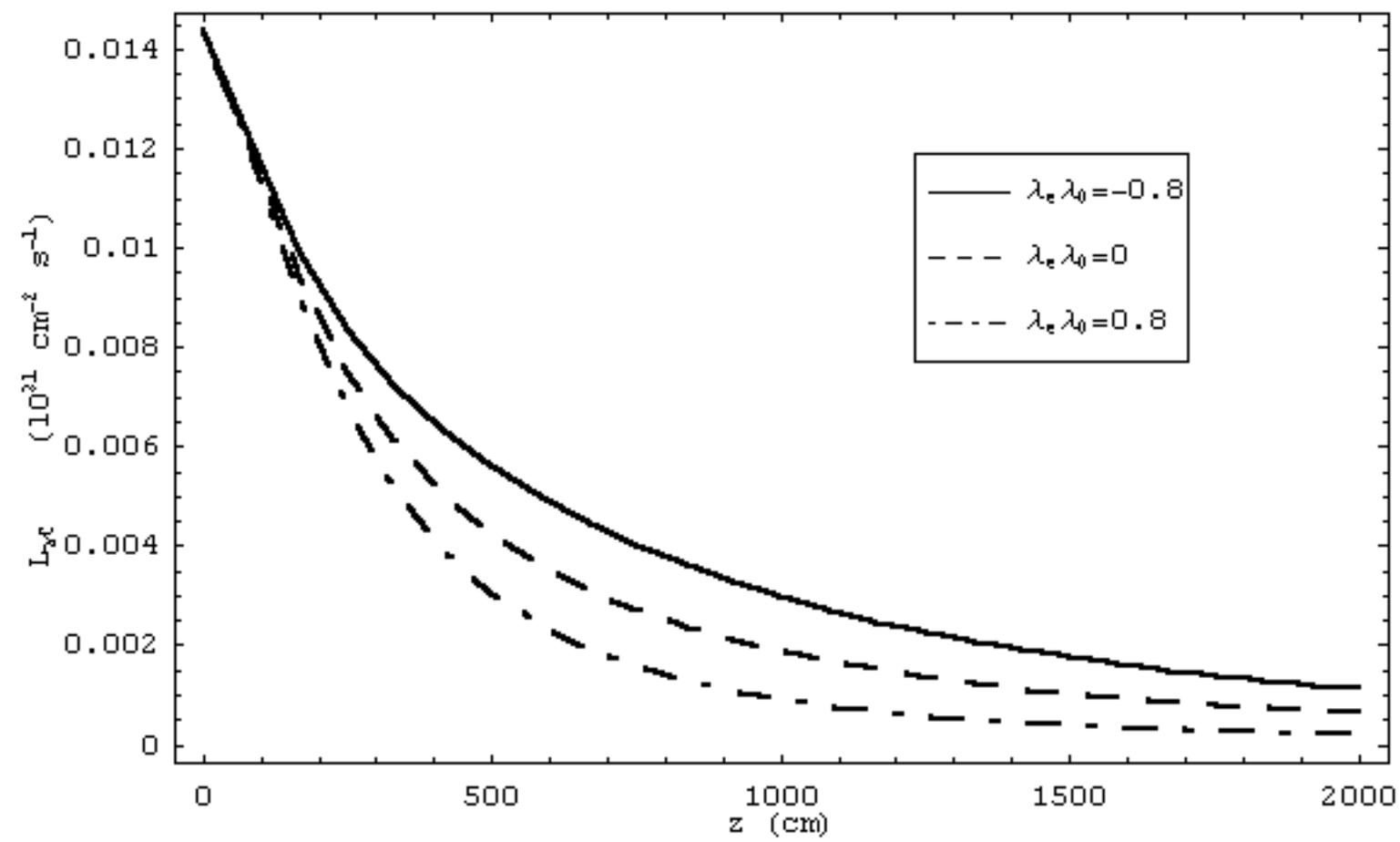
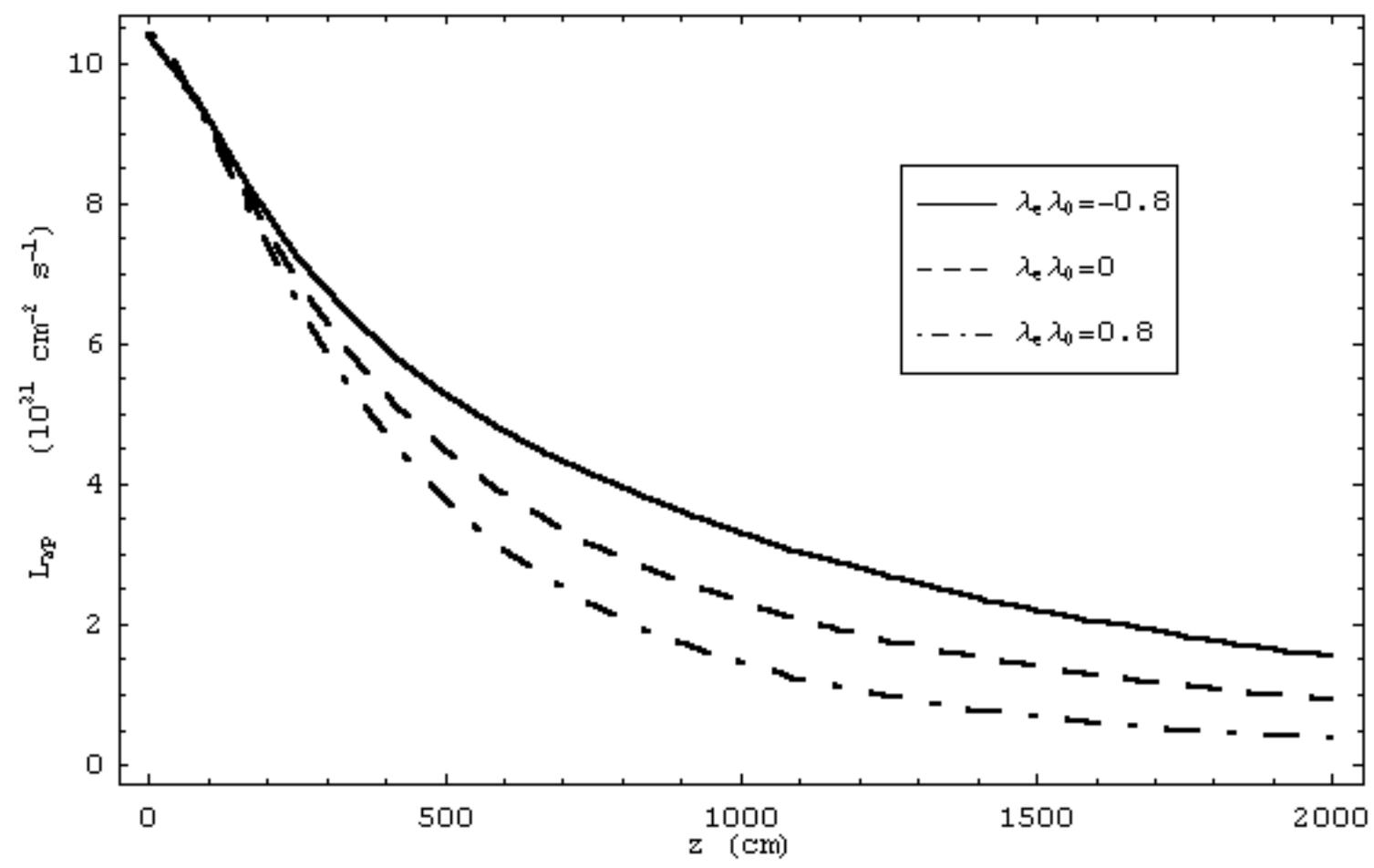
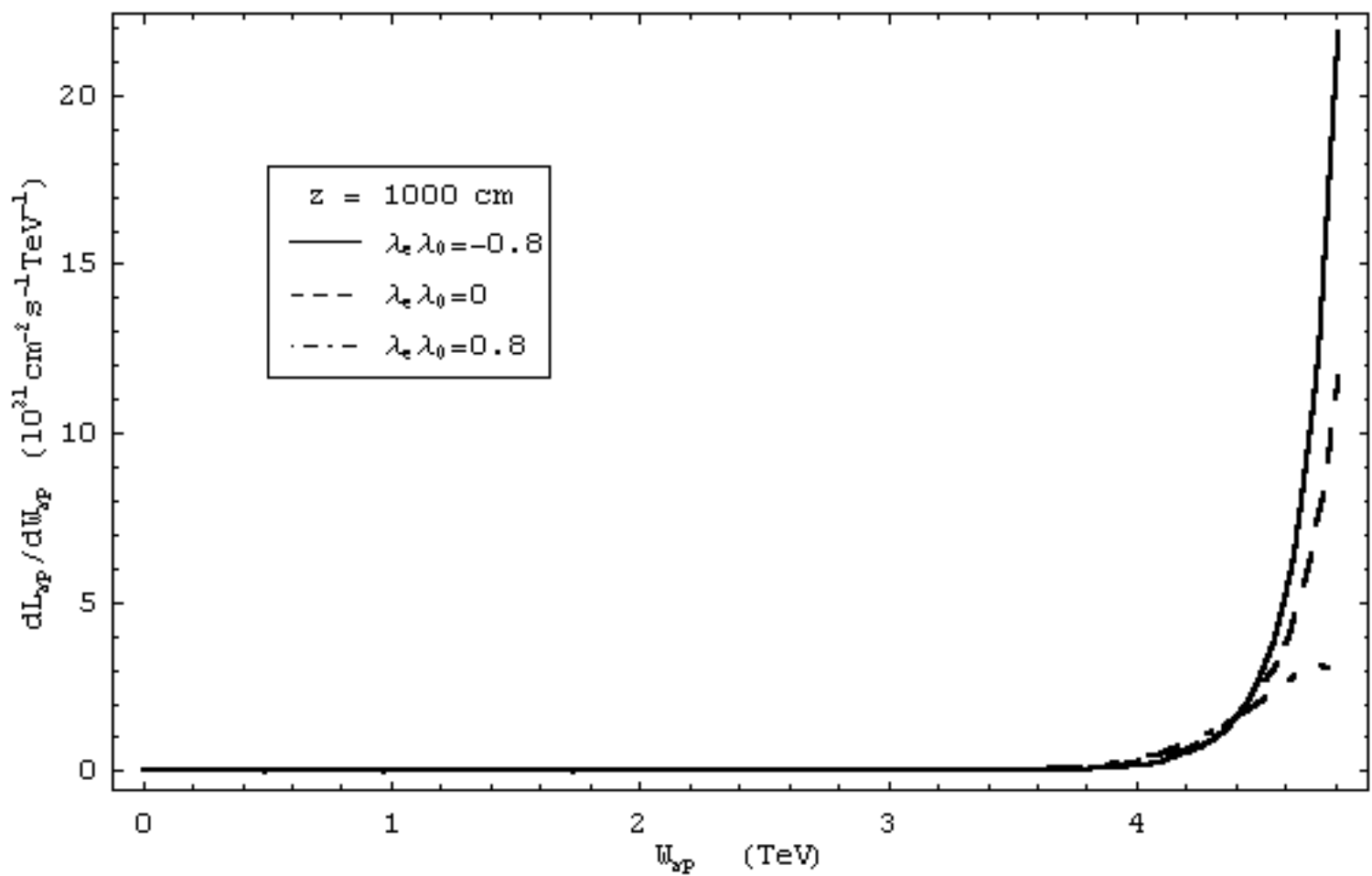
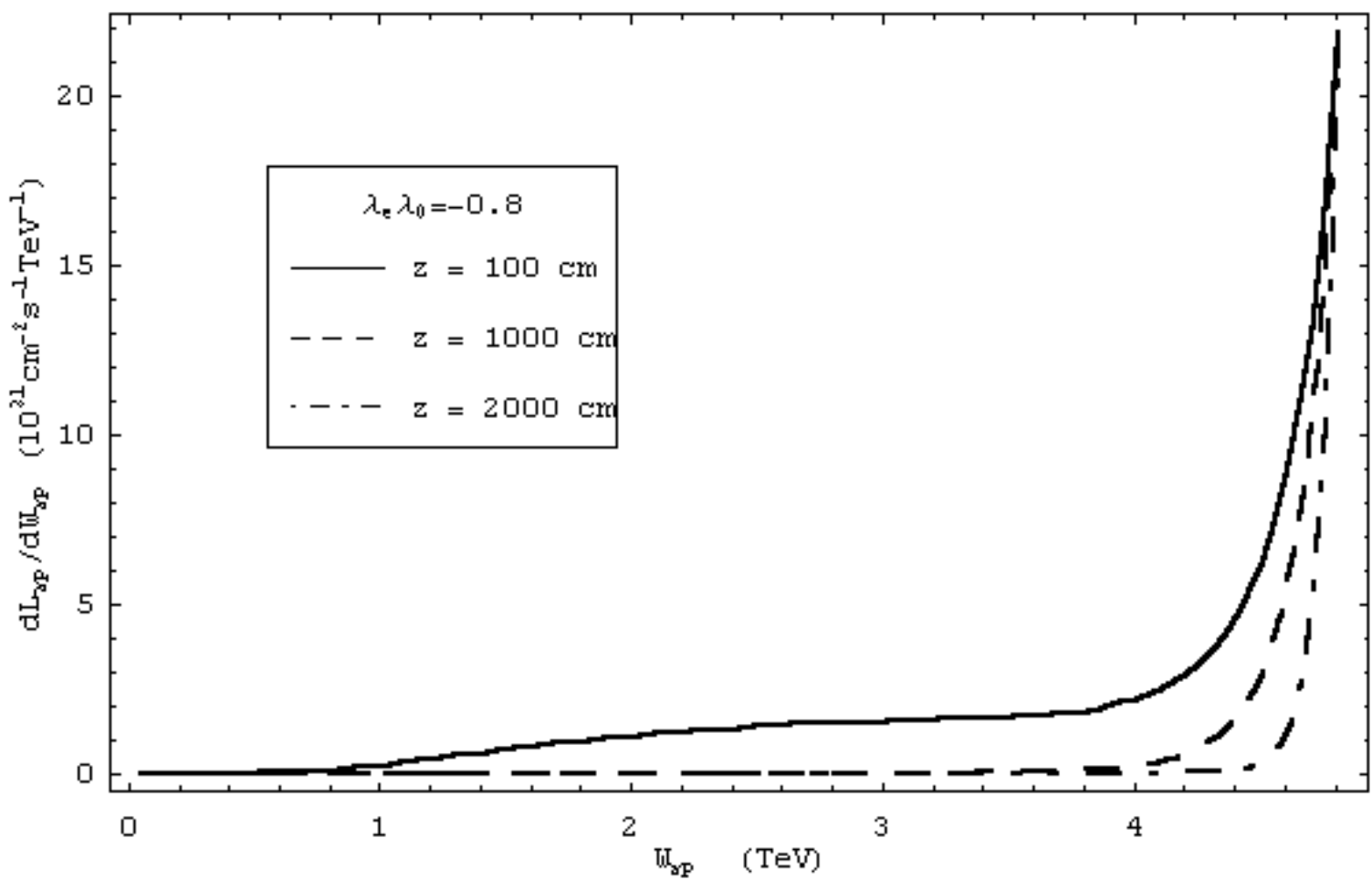
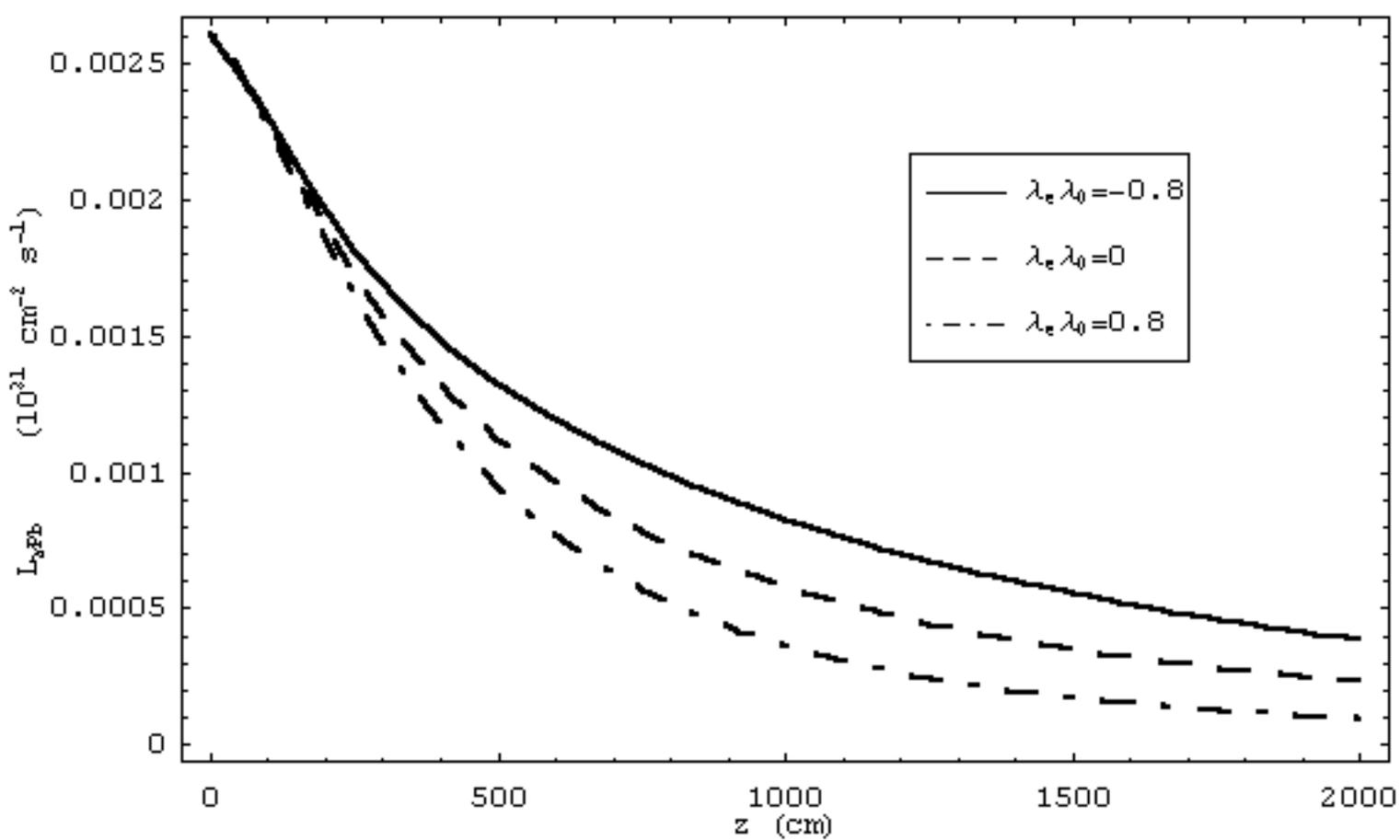
\end{document}